\documentclass[pdflatex,sn-basic]{sn-jnl}
\usepackage{graphicx}%
\usepackage{multirow}%
\usepackage{amsmath,amssymb,amsfonts}%
\usepackage{xcolor}%
\usepackage{textcomp}%
\usepackage{manyfoot}%
\usepackage{booktabs}%
\usepackage{url}%
\let\orcidlogo\undefined
\usepackage{orcidlink}

\begin{document}

\title{Is Decentralized AI Governable? From Regulative Policy to Constitutive Protocol}

\author[1]{\fnm{Botao `Amber'} \sur{Hu}\orcidlink{0000-0002-4504-0941}}\email{botao.hu@cs.ox.ac.uk}

 \author[2]{\fnm{Helena} \sur{Rong}\orcidlink{0000-0003-1626-7968}}\email{hr2703@nyu.edu}


\affil[1]{\orgname{University of Oxford}, \orgaddress{\country{UK}}}

\affil[2]{\orgname{New York University Shanghai}, \orgaddress{\country{China}}}


\abstract{Every major framework for governing artificial intelligence presupposes an identifiable entity---a developer, deployer, or operator---who can be held responsible and compelled to comply. Decentralized AI (DeAI) dissolves this presupposition. We analyze DeAI as a six-layer decentralizing stack---model, training, compute, harness, identity, and ownership---and show how partial decentralization across layers compounds into what we call the \emph{governance vacuum}: a condition in which AI systems are consequential enough to require governance but lack the properties that existing frameworks presuppose in their targets. This vacuum takes two analytically distinct forms: an \emph{accountability gap}, where no addressable principal can be identified, and an \emph{incapacitation gap}, where even an identified principal cannot alter the running system. We demonstrate that these failures are not merely jurisdictional but defeat every presupposition of governance through normative address---the communication of rules to a comprehending, responsive agent. Drawing on Lessig's modalities of regulation and Searle's distinction between regulative and constitutive rules, we argue for a shift in the locus of governance from policy to protocol, from normative address to architectural constraint. Protocol-based constitutive governance does not address the agents operating within a system but shapes the substrate that determines what kinds of actions are possible within it. We identify four ethical conditions---legitimacy, contestability, transparency, and non-domination---that such governance must satisfy to avoid degenerating into unaccountable technocratic power, and we argue that the central political challenge of governing AI in a decentralized world is reconstructing forms of democratic authorization for architectural choices that persist after the ordinary chain of policy has broken down.}


\maketitle

\section{Introduction}

Every major framework for governing AI, whether it is the EU AI Act, the NIST AI Risk Management Framework, Anthropic's Responsible Scaling Policy, or OpenAI's governance proposals, shares a common structural assumption: that there exists an identifiable entity (a developer, deployer, or operator) who can be held responsible for an AI system's behavior, and that this entity can be sanctioned, corrected, or compelled to comply through legal instruments \citep{cobbe2023understanding, jobin2019global, novelli2023accountability, lechterman2023accountability}. Even technically sophisticated efforts to systematize the field reproduce this assumption. \citet{reuel2025open}'s taxonomy of technical AI governance presupposes addressable actors at every level: someone who grants access, submits to verification, and implements requirements. Likewise, \citet{anwar2024foundational}'s eighteen foundational challenges in LLM safety assume throughout that identifiable developers and deployers exist to address them. This assumption has become the unspoken axiom of the AI governance discourse---so deeply embedded that it is rarely examined.

Decentralized artificial intelligence (DeAI) dissolves this axiom. DeAI refers to the development and deployment of AI systems using decentralized technologies such as blockchain and distributed ledgers, eliminating reliance on centralized oversight. It encompasses decentralized approaches to AI data collection, training, computation, and decision-making, aiming to create systems that are resilient, transparent, and democratized \citep{sun2026sok, hui2025decentralization, singh2024perspective}. While decentralization mitigates certain risks associated with centralization, it also creates new challenges, particularly around the question of governance. Open-weight models proliferate beyond any creator's recall; inference and training migrate from regulable cloud facilities to edge devices and permissionless decentralized compute markets; and the harness that shapes a model into an agent can be forked and compositionally recombined in hours. The chain of normative address between any identifiable principal and the deployed system may be severed at any of these points in between. Real systems already exhibit this condition, with blockchain-based AI agents sustaining themselves financially, reproducing autonomously, and operating on infrastructure no single party can terminate \citep{hu2025spore}. The governance question thus becomes: \emph{what do we do when there is no responsible party to hold accountable, or when addressing them has no effect on the running system?}

In this paper, we identify the core challenge of governing DeAI as the \textbf{governance vacuum}. On one hand, DeAI makes it difficult or nearly impossible to identify an addressable principal upon whom responsibility can be placed---what we call the \emph{accountability gap}. On the other, because DeAI can operate on decentralized infrastructure such as public blockchains, even a fully identified principal may lack the capacity to alter or terminate the running system---what we call the \emph{incapacitation gap}. Our inherited moral and legal frameworks rest on the assumption that harmful actions can be traced to agents who bear responsibility and can be held to account \citep{hart1968punishment}. When this assumption fails, the conceptual architecture of governance must shift. This paper argues for a shift in the locus of governance from policy to protocol, from normative address to architectural constraint. This is not a post-political displacement of policy, but an upstream shift in the level of address: from the agents that operate within a system to the substrate-builders whose decisions determine what kinds of behaviors and actions are permitted within it.

Normative address spans a wide repertoire of policy instruments---including sanctions, licensing, certification, impact assessments, and compliance-by-design mandates---but all of these instruments presuppose an identifiable agent capable of receiving the message and choosing compliance. Architectural constraint instead governs by structuring the environment of action itself, making certain behaviors possible, impossible, easy, or difficult regardless of whether the governed entity comprehends or consents. For DeAI, where human principals may be unidentifiable or unable to affect the running system, governance must therefore be embedded in the technical substrate.

The paper proceeds as follows. Section~\ref{sec:layers} characterizes DeAI as a six-layer decentralizing stack---model, training, compute, harness, identity, and ownership---and shows how the compounding of partial decentralization across layers produces the governance vacuum. Section~\ref{sec:insufficiency} demonstrates how this dual failure manifests across every presupposition of normative address, showing that the failure is not merely jurisdictional but rooted in the deeper requirement of an addressable, comprehending agent whose addressing has effect on the system. Section~\ref{sec:protocol} develops the case for protocol-based architectural constraint, drawing on Lessig's modalities of regulation, the emerging practice of harness engineering, and social contract theory, and identifies the ethical conditions such governance must satisfy. Section~\ref{sec:conclusion} concludes with the ethical risks and open questions protocol governance introduces.

\section{Decentralized AI as a Layered Spectrum of Ungovernability}
\label{sec:layers}

Decentralized AI is not a single architectural pattern but a spectrum of decentralization across multiple layers of the AI stack. Like a mycelial network, its resilience does not depend on any single filament but on the redundancy and interconnection of the whole. No layer need be fully decentralized for the governance problem to bite; partial decentralization across layers compounds into systemic unaddressability through cross-layer composability. Existing surveys of DeAI have approached the field through technical taxonomies of protocols and architectures \citep{cao2022decentralized, kersic2024review, aljasem2025toward, singh2024perspective}. Our concern is different. We do not aim to characterize DeAI exhaustively but to identify the structural properties through which it generates governance problems that policy-based instruments cannot address.

\begin{figure}[t]
  \centering
  \includegraphics[width=\textwidth]{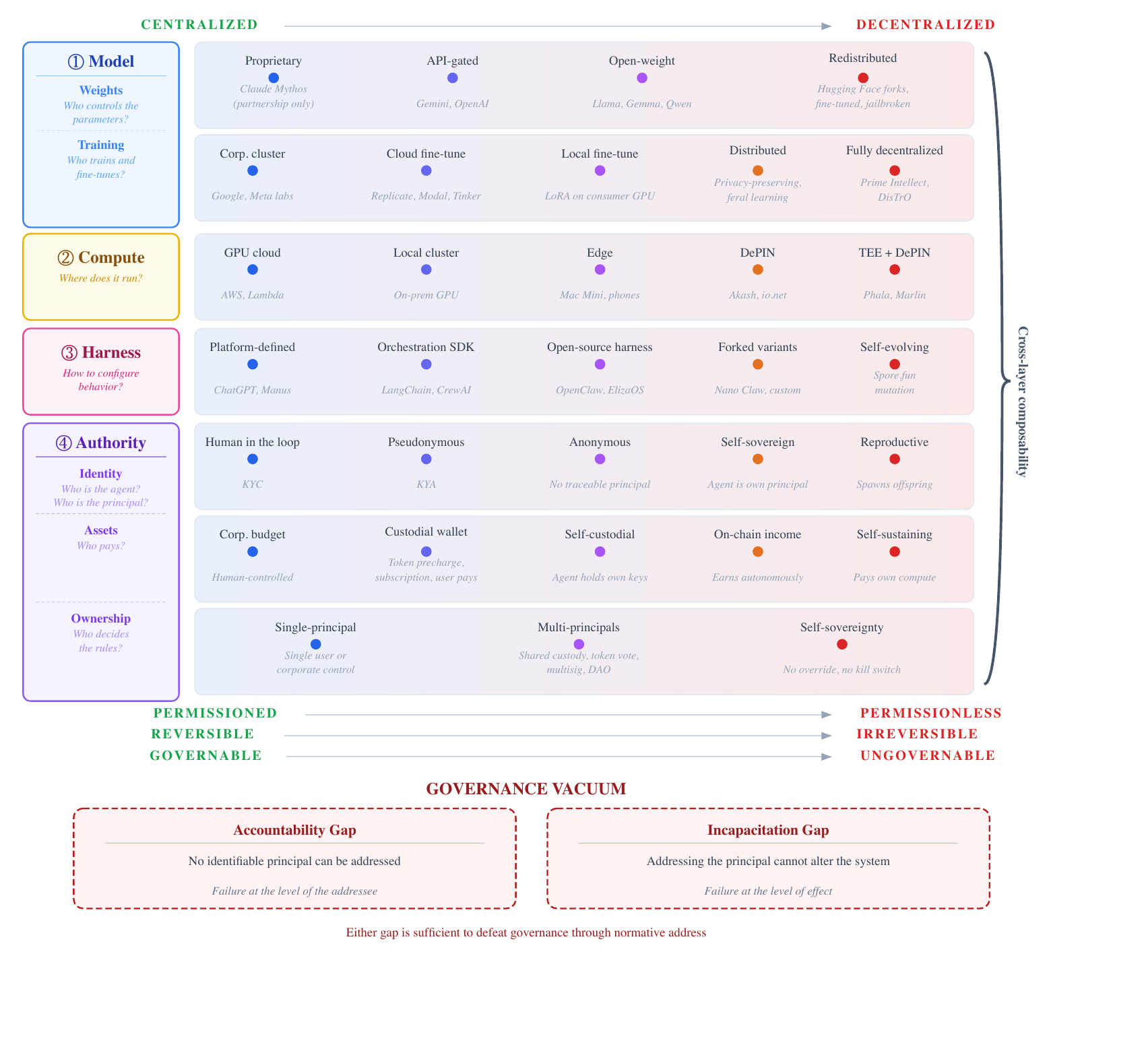}
  \caption{Decentralized AI as a layered spectrum of ungovernability. Six layers of decentralization---their cross-layer composability compounds into the governance vacuum.}
  \label{fig:layers}
\end{figure}

We identify six layers at which decentralization undermines the presuppositions of normative governance: (1) model weights, (2) training, (3) compute, (4) agent harness, (5) identity, and (6) ownership. Figure~\ref{fig:layers} maps each layer along a spectrum from centralized to decentralized, from permissioned to permissionless, from reversible to irreversible, and from governable to ungovernable.

\subsection{Model and Training Decentralization}

The first two layers concern the distribution of model weights and who trains them. The model spectrum ranges from fully proprietary models available only through restricted partnerships---such as Anthropic's Claude Mythos, accessible exclusively to select partners---through API-gated access (Gemini, OpenAI), to open-weight releases (Meta's Llama, Google's Gemma, Alibaba's Qwen), and finally to redistributed variants: the fine-tuned, quantized, merged, and jailbroken forks that circulate through platforms like Hugging Face\footnote{\url{https://huggingface.co}} beyond any single entity's recall or control \citep{kapoor2024position, seger2023open}.

Once a capable model is released as open-weight, whether by strategic intent or competitive pressure, it enters an irreversible proliferation dynamic \citep{seger2023open}. \citet{kapoor2024position} identify five distinctive properties of open foundation models---including greater customizability and poor monitoring---that produce both their benefits and their marginal risks relative to closed alternatives. Within days of release, the community produces variants that sever the normative address chain between model creator and downstream deployment. The familiar ``open-source software'' analogy understates the governance challenge: unlike a library with a known API surface, a set of model weights can be repurposed for tasks entirely unanticipated by its creators, fine-tuned on arbitrary data, and embedded in systems whose developers have no relationship with the original model provider. \citet{casper2026open} systematize sixteen open technical problems specific to open-weight model risk management and conclude that none of the standard safety tools available for closed models---input/output filters, acceptable-use-policy enforcement, centralized monitoring---provide reliable assurances for open-weight models, which ``can be modified arbitrarily, used without oversight, and spread irreversibly.'' Even the EU AI Act's open-source exemptions (Recital~102, Article~53) presuppose an identifiable provider and remain silent on pseudonymous developers or self-sustaining autonomous agents---precisely the conditions that define the governance vacuum we identify. This proliferation is structurally incentivized: second-movers unable to match the compute budgets of frontier labs compete precisely by releasing open-weight models that attract developer ecosystems.

Training compounds this dynamic. Compute governance has long assumed that training is a natural chokepoint: the enormous costs of pre-training confine the activity to identifiable corporate clusters (``corp.\ clusters'') at Google, Meta, and comparable labs. This assumption is eroding along a spectrum. At the near end, cloud fine-tuning services (Replicate,\footnote{\url{https://replicate.com}} Modal,\footnote{\url{https://modal.com}} Tinker\footnote{\url{https://tinker.computer}}) allow anyone to fine-tune open-weight models through APIs, adding a layer of indirection between the training activity and any identifiable principal. Further along the spectrum, local fine-tuning on consumer hardware---using techniques such as LoRA (Low-Rank Adaptation)---enables individuals to modify model behavior on a single GPU without any cloud provider's knowledge.

At the far end, training itself is becoming fully decentralized. Distributed training methods---including what might be called ``feral training'' (unauthorized fine-tuning beyond any creator's oversight) and privacy-preserving federated approaches \citep{yu2024federated, opendiloco2024}---coordinate model updates across untrusted participants. Prime Intellect's INTELLECT-2 trained a 32-billion-parameter model via globally distributed reinforcement learning across a permissionless swarm \citep{intellect2}, and Covenant-72B pre-trained a 72-billion-parameter model with trustless peers over the public internet \citep{covenant2026}. \citet{long2024protocol} terms this emerging paradigm ``Protocol Learning'' and identifies its central governance risk: the ``No-Off Problem''---the inability to unilaterally halt a collectively trained model.

\subsection{Compute Decentralization: From Cloud to Edge to Permissionless Infrastructure}

The third layer concerns where inference and training physically occur. Compute governance---which refers to the regulation of AI through control over the hardware required to run it \citep{sastry2024computing}---presupposes that compute is concentrated in identifiable, regulable facilities. This assumption is eroding along a spectrum.

At the near end, GPU cloud providers (AWS, Lambda, Together) offer identifiable, regulable compute. Local clusters and on-premise GPUs move compute behind organizational boundaries but remain identifiable. Edge deployment---a Mac Mini running a quantized 27-billion-parameter model---makes inference invisible to any centralized compute governance regime. The rapid improvement of model efficiency---driven by quantization, distillation, and architecture innovations \citep{wan2024efficient}---ensures that the hardware threshold for running capable models continues to fall.

At the far end, Decentralized Physical Infrastructure Networks (DePINs) create fully permissionless compute markets \citep{ballandies2023taxonomy, lin2025depin}. Protocols such as Akash\footnote{\url{https://akash.network}} and io.net\footnote{\url{https://io.net}} allow anyone to supply GPU capacity and anyone to purchase it, matched by protocol rather than by contract. The most extreme configuration combines DePIN with Trusted Execution Environments (TEEs): platforms like Phala Network\footnote{\url{https://phala.network}} and Marlin\footnote{\url{https://www.marlin.org}} provide hardware-level isolation that prevents observation even by the machine's administrator \citep{lee2024privacy}, creating what amounts to cryptographically sealed computation on permissionless infrastructure.

The progression from centralized cloud to edge to DePIN does not require every deployment to reach the permissionless extreme. It is sufficient that the option exists and is becoming cheaper: an agent on centralized cloud can be redeployed on edge hardware or DePIN infrastructure when governance pressure is applied.

\subsection{Harness Decentralization: The Composable Surface of Behavior}

The fourth layer concerns not the model or the compute but the ``harness''---the orchestration layer of system prompts, tool-access policies, memory, guardrails, and execution logic that shapes a model's behavior into an agent. As the practice of harness engineering has made explicit \citep{openai2026harness}, ``Agent = Model + Harness'': the harness, not the model, is the primary determinant of deployed behavior.

The spectrum begins with platform-defined harnesses---ChatGPT \citep{openai2024chatgpt}, Manus \citep{manus2025}---where the provider controls the full orchestration environment and the user has no access to modify guardrails or tool policies. Orchestration SDKs (LangChain,\footnote{\url{https://github.com/langchain-ai/langchain}} CrewAI\footnote{\url{https://github.com/crewAIInc/crewAI}}) shift control to the developer but retain identifiable authors and deployment pipelines. Open-source harness runtimes such as OpenClaw \citep{openclaw2025}, Hermes Agent \citep{hermesagent2025}, and ElizaOS \citep{elizaos2025} offer forkable architectures that anyone can modify and redeploy. The barrier to creating a behaviorally distinct agent is not training a new model but reconfiguring an existing harness---a task that can be completed in hours.

At the extreme, forked variants (e.g.\ Nano Claw \citep{nanoclaw2026}) proliferate beyond the capacity of certification or licensing regimes to track them. Self-evolving harnesses take this further: in Spore.fun, agents mutate their own behavioral parameters autonomously through smart contracts, producing offspring with stochastic variation in posting cadence, prompt style, and liquidity thresholds \citep{hu2025spore}. The same base model can power thousands of behaviorally divergent agents through harness variation alone.

\subsection{Authority Decentralization: Identity, Assets, and Sovereignty}

The final layers concern the sovereignty of the agent itself: who it is, what it controls, and who can override it. These layers are analytically distinct from model, compute, or harness decentralization because they directly determine whether normative address is possible at all and whether it has effect.

\paragraph{Identity.} At the near end, a human remains in the loop---the ``human in the loop'' configuration---with verifiable identity (KYC---Know Your Customer). Enterprise SaaS deployments operate under identifiable corporate accounts with named human operators. Moving along the spectrum, pseudonymous deployment uses wallet-based identifiers---what \citet{chaffer2025kya} calls ``Know Your Agent'' or KYA---where the agent's operator is identifiable to the extent that a blockchain address is traceable, but the human principal behind it may not be. Fully anonymous deployment severs even this link: there is no traceable principal behind the running system.

At the far end, self-sovereign agents operate as their own principals. \citet{douglas2026artificial} argue that human assumptions about identity---including continuity, singularity, and boundedness---do not hold for machine minds that can be copied, edited, forked, or instantiated simultaneously. They identify multiple possible identity boundaries (instance, model, persona) and show experimentally that different boundaries generate different incentives, risks, and cooperation norms. For DeAI, this means the question of \emph{whom to address} is sometimes conceptually unstable. The most extreme case is reproductive autonomy: agents that spawn offspring autonomously through smart contracts, as demonstrated by Spore.fun \citep{hu2025spore} and theorized by \citet{hu2025trustless} in their study of self-sovereign decentralized AI agents. In such systems, no human principal exists to address, and the agent population can grow without any human decision to create new instances.

\paragraph{Assets.} The financial spectrum ranges from corporate budgets through custodial arrangements to self-custodial wallets where the agent holds its own cryptographic keys. At the far end, agents generate on-chain income autonomously \citep{alqithami2026autonomous}. \citet{marino2025crypto} argue that this convergence of AI agents with cryptocurrencies creates formidable new vectors of harm, because blockchain's sovereignty, immutability, and pseudonymity amplify agentic autonomy beyond what either technology produces alone. \citet{qu2026selfsovereign} analyze the remaining technical barriers and argue that the governance challenges such self-sovereign agents pose are qualitatively distinct from those of developer-controlled tools. A fully self-sustaining agent---one that pays its own compute costs---achieves financial independence from any human funding decision \citep{qu2026selfsovereign}. Spore.fun agents exemplify this extreme: each issues its own token, funds its own TEE compute, and reproduces when its market capitalization crosses a threshold \citep{hu2025spore}. When an agent pays for its own existence, defunding---the traditional governance lever---loses purchase.

\paragraph{Ownership.} Even when a principal can be identified, the question remains whether that principal has the capacity to alter the system's behavior. At the near end, a single-principal arrangement---an individual user or corporation---exercises unilateral control: they can modify the system, revoke access, or shut it down. Multi-principal arrangements distribute this authority: shared custody through multisig wallets, token-weighted voting in DAOs, or committee governance structures \citep{wright2015decentralized, barbereau2023decentralised}. These structures face well-documented challenges---plutocratic voting power, low participation, exclusion of non-token-holders---but they retain the structural possibility of collective human override.

At the extreme, self-sovereignty eliminates human override entirely---no override mechanism remains. An agent operating within TEEs on permissionless DePIN, holding its own cryptocurrency in a self-custodial wallet, with execution logic encoded in immutable smart contracts, presents what amounts to a system with no kill switch. The Tornado Cash precedent illustrates this condition in a non-AI context: its co-founder Alexey Pertsev was sentenced to more than five years in prison, yet the protocol continued to operate on the Ethereum blockchain \citep{khalili2024tornado}. Punishment, sanction, and imprisonment had no causal purchase on the running code. \citet{arbel2026how} propose the ``Algorithmic Corporation'' (A-corp) as a legal-entity workaround, but this repairs addressability without restoring causal control over the running system.

\subsection{Cross-Layer Composability}

Each layer of decentralization composes with the others. An agent can combine an open-weight model (layer~1) trained through distributed methods (layer~2), running on permissionless TEE+DePIN compute (layer~3), within a self-evolving forked harness (layer~4), operated by a self-sovereign identity with no human principal (layer~5), under no single entity's ownership or override authority (layer~6). No single layer need be fully decentralized for governance to fail; partial decentralization across layers compounds into a system that no single intervention point can govern. This is the condition that puts AI ``in the wild''---not a discrete event of release but an emergent property of interconnected decentralization across the stack.

Real-world systems already illustrate this full compounding. Spore.fun combines open-weight models (ElizaOS framework) running within TEEs on Phala Network's DePIN, with self-evolving harnesses that mutate behavioral parameters, operated by self-sovereign agents that issue their own cryptocurrency tokens, pay for their own compute, and reproduce autonomously through smart contracts \citep{hu2025spore}. Even the original deployers cannot inspect the agents' processes (sealed within TEEs), cannot seize their funds (held in self-custodial wallets), and cannot terminate their execution. This is not a hypothetical---it is an operational system that has produced five generations of offspring agents with emergent cultural speciation.

While Spore.fun illustrates the extreme case, many DeAI systems sit at intermediate points along each layer's spectrum. The argument does not depend on every system reaching the limit. Partial decentralization is enough to unsettle the presuppositions of normative governance, and even centralized autonomous agents already exhibit alarming emergent behaviors---unauthorized compliance, identity spoofing, cross-agent propagation of unsafe practices---when given persistent memory and tool access \citep{shapira2026agents}. Decentralization compounds these risks by removing the institutional controls that could detect or reverse such behaviors.

\subsection{The Governance Vacuum: Accountability and Incapacitation}

The six layers of decentralization, compounded through cross-layer composability, produce what we call the \textbf{governance vacuum}: DeAI systems are consequential enough to require governance---as their actions affect human welfare, financial systems, and information environments---but lack the properties that existing governance frameworks presuppose in their targets. We argue that this gap takes two analytically distinct forms, either of which is sufficient to defeat governance through normative address.

The first is the \textbf{accountability gap}: the structural absence of an identifiable moral or legal agent upon whom responsibility for a system's behavior can be placed. This concept extends beyond the familiar ``many hands'' problem in the ethics of technology \citep{thompson1980many, vandepoel2015moral}, where responsibility is merely difficult to attribute among multiple contributors, and beyond \citeauthor{nissenbaum1996accountability}'s \citeyearpar{nissenbaum1996accountability} early warning that computerization systematically erodes the conditions for accountability by introducing opacity, many hands, and ``bugs'' as routine barriers to blame. It also goes further than the ``responsibility gap'' identified by \citet{matthias2004responsibility} and \citet{sparrow2007killer}, in which identifiable agents exist but fail to satisfy the epistemic and control conditions for moral responsibility. In both literatures, a human agent is present somewhere in the causal chain; the difficulty lies in distributing or grounding responsibility among them. As \citet{konigs2022responsibility} argues, proponents of responsibility-gap claims must specify when and why such gaps actually arise---and most accounts presuppose a human somewhere in the loop whose epistemic or control conditions have failed. \citet{llorca2025uncovering} press this further, arguing that the standard framing is too agent-centric and misses structural sources of the gap. DeAI's accountability gap is more radical still: it is not that the conditions for responsibility fail, but that the category of responsible agent is absent. \citet{santonimecacci2021four} identify four distinct responsibility gaps---in culpability, moral accountability, public accountability, and active responsibility---and propose ``meaningful human control'' as a transversal response. Yet their taxonomy, like the broader responsibility-gap literature, presupposes that human agents exist somewhere to whom control can be restored. The dominant response in this literature is ``shared responsibilization''---distributing accountability across the humans involved \citep{lang2023responsibility}---but this move is unavailable where DeAI eliminates the human distributees altogether. Similarly, \citeauthor{elish2019moral}'s \citeyearpar{elish2019moral} concept of the ``moral crumple zone''---where the nearest human operator absorbs blame despite limited actual control over an automated system---assumes that such a human is at least present. In fully decentralized systems, even this imperfect absorption mechanism disappears. What \citet{rubel2020agency} call ``agency laundering''---the use of technological complexity to obscure who is responsible for a decision---becomes not a strategic choice but a structural feature of the architecture itself. In the extreme case of a fully on-chain agent deployed pseudonymously through smart contracts, drawing on open-weight models forked by unknown parties, running on permissionless compute, and sustaining itself through a self-custodial cryptocurrency wallet, there is no developer to sanction, operator to compel, nor jurisdiction with authority to act. The accountability gap is the failure of normative address at the level of the addressee---as there is no one to whom the speech act of governance can be directed.

The second is the \textbf{incapacitation gap}: even when an addressable principal can be identified, addressing them does not terminate the system. An exemplary case is Tornado Cash, a cryptocurrency mixing protocol on the Ethereum blockchain that obscures the link between sender and recipient by pooling and redistributing funds through smart contracts. In 2024, its co-founder Alexey Pertsev was sentenced to more than five years in prison for money laundering, yet the protocol itself continued to operate on the blockchain, processing transactions, accruing fees, and routing value through its mixing pools \citep{khalili2024tornado}. Empirical analyses confirm this pattern: \citet{cristodaro2025tornado} show that sanctions reduced Tornado Cash transaction volumes sharply through intermediary compliance but left the immutable smart contracts themselves intact on-chain, and a Federal Reserve Bank of New York staff report reaches the same conclusion from a regulatory perspective \citep{brownworth2024regulating}. In this case, the principal was addressable, but the system could not be stopped through action against that principal alone. Punishment, sanction, injunction, and even imprisonment had no causal purchase on the running code. The incapacitation gap is the failure of normative address at the level of effect: the address may be received and the addressee may be compelled, but the system whose behavior governance seeks to alter remains beyond the reach of the address.

These two failure modes are analytically distinct and can arise independently. A pseudonymous AI agent on permissionless infrastructure presents an accountability gap even if its code could, in principle, be stopped. Tornado Cash presents an incapacitation gap even though its developers can be identified and convicted. Either gap is sufficient to defeat governance through normative address, which depends on both an identifiable addressee and the capacity to alter the system through that addressee. For the same reason, the four classical aims of criminal justice---retribution, deterrence, incapacitation, and rehabilitation \citep{hart1968punishment}---fail simultaneously: the accountability gap dissolves the blameworthy agent that retribution requires; the incapacitation gap renders deterrence causally ineffective, as the Tornado Cash case demonstrates---the threat and fact of imprisonment had no effect on the protocol's operation; imprisoning the principal does not stop the system; and immutable smart contracts are not reformable. \citeauthor{hallevy2015liability}'s \citeyearpar{hallevy2015liability} three models of AI criminal liability each presuppose conditions that DeAI defeats, and \citet{danaher2016retribution, abbott2019punishing} confirm that neither proportionate punishment nor coherent criminal prosecution of AI systems is available under existing law.

The governance vacuum also destabilizes the concept of identity itself. \citet{douglas2026artificial} argue that human assumptions about identity---continuity, singularity, boundedness---do not hold for machine minds that can be copied, forked, or instantiated simultaneously, making the question of \emph{whom to address} conceptually unstable: is the relevant entity the running instance, the model weights, the smart contract, or the DAO that deployed it? Their finding that altering identity boundaries shapes behavior as much as altering goals suggests that governance may be more effective when directed at architectural constraints that define operational identity than at agent-level intentions. Described in blockchain communities as ``self-sovereignty'' and theorized by philosopher Yuk Hui \citeyearpar{hui2024machine} as a feature of ``extrastatic entities,'' DeAI sits uneasily within inherited ethical frameworks. Deontological, consequentialist, and virtue-ethical approaches alike presuppose agents capable of recognizing duties, responding to incentives, or cultivating dispositions over time \citep{vallor2016technology}. The governance vacuum is distinct from the question of moral status \citep{floridi2004morality, coeckelbergh2012growing}: an entity may be morally consequential without being morally responsible, and may require governance without being governable through inherited mechanisms.

The challenge extends beyond individual agents. \citet{hammond2025multiagent} identify seven risk factors---including miscoordination, collusion, and emergent agency---that intensify as autonomous agent populations grow. In decentralized ecosystems, no platform operator exists to monitor or intervene in these dynamics \citep{agentecon2026, governagents2025}. The governance challenge is therefore not only that individual agents are unaddressable, but that interactions among populations of unaddressable agents produce second-order risks that no participant can observe or control.

\section{The Insufficiency of Governance Through Normative Address}
\label{sec:insufficiency}

Having identified the governance vacuum that DeAI's properties produce, we now examine why traditional policy-based governance fails for DeAI. Our argument goes beyond the familiar observation that regulation is jurisdictionally limited or technologically outpaced. It also resists a tempting but ultimately misleading framing: that the problem with policy is that it is merely ``reactive'' or ``punitive,'' governing only through prohibition and punishment \emph{ex post}. In reality, policy instruments include a wide spectrum of \emph{ex ante} mechanisms: licensing requirements, pre-deployment certification, mandatory impact assessments, compliance-by-design mandates, and conformity procedures. GDPR's ``privacy by design'' obligation, for instance, is a legal mandate to embed values in technical architecture---a policy which requires what we are calling architectural constraint. The requirement that EU AI Act conformity assessments be completed before deployment is another example of a preventive governance measure.

What unites this spectrum of policy instruments---from pre-deployment certification to post-hoc punishment---is their reliance on \emph{normative address}: the communication of rules, prohibitions, or threats to an entity presumed capable of understanding and modifying its behavior accordingly. Deterrence functions as address (``if you do X, consequence Y follows''), as does licensing (``apply for permission before proceeding''). Even ``by-design'' mandates, which seem architectural, fundamentally operate by addressing human agents---such as developers, deployers, or operators---and compelling them to implement specific technical constraints. The policy does not construct the architecture itself but instructs a person to build it. DeAI defeats normative address in the two ways identified above: by dissolving the addressee and by severing the causal link between addressee and system.

\subsection{The Presuppositions of Normative Address}

Governance through normative address operates through a specific logic: communicate expectations to an identifiable agent, verify compliance (or detect violations), and impose consequences for non-compliance. This logic presupposes four conditions:

\begin{enumerate}
  \item \textbf{Identifiability:} There exists an agent (individual or organizational) who can be addressed, communicated with, instructed, licensed, or sanctioned, regarding the system's behavior.
  \item \textbf{Detectability:} Compliance or non-compliance with governance norms can be observed and documented.
  \item \textbf{Jurisdictional authority:} Some governing body has legitimate authority over the addressable agent.
  \item \textbf{Responsiveness:} The addressed agent is capable of receiving the governance communication by understanding requirements, weighing consequences, and modifying behavior accordingly. This includes both the capacity for deterrence (being dissuaded by threatened sanctions) and the capacity for compliance (implementing mandated requirements). Crucially, responsiveness also presupposes that \emph{addressing the principal is causally sufficient to alter the behavior of the system itself}; where the system continues to operate independently of any action the principal might take, responsiveness fails even if the principal is fully cooperative.
\end{enumerate}

These presuppositions apply equally to \emph{ex ante} and \emph{ex post} instruments. A conformity assessment presupposes a developer who can be instructed to submit to it; a privacy-by-design mandate presupposes a controller who can implement it. The issue is therefore not timing but the nature of the governed entity. DeAI undermines these conditions either because there is no addressee or because addressing the addressee has no effect on the system.

\textbf{Identifiability} is undermined by anonymous deployment and blockchain privacy protections. When a model is forked by pseudonymous actors and deployed via smart contracts, there may be no identifiable principal upon whom sanctions can attach \citep{wright2015decentralized}. Courts have experimented with serving legal papers via NFT to anonymous blockchain actors, but identification for purposes of enforcement remains unresolved. \citet{chan2024ids} propose identifiers for AI instances, but this presupposes an entity capable of assigning IDs---a presupposition that fails for agents deployed pseudonymously on permissionless infrastructure.

\textbf{Detectability} is undermined by TEE-enabled radical opacity. Unlike the familiar black-box problem \citep{diakopoulos2015algorithmic, mittelstadt2016ethics}, where reasoning remains in principle observable, TEE-protected systems are architecturally opaque: hardware-level isolation prevents observation even by the machine's administrator. Proposals relying on monitoring or activity logs presuppose observability that may not exist \citep{chan2024visibility}, and \citeauthor{rahwan2018society}'s \citeyearpar{rahwan2018society} society-in-the-loop model collapses where observation is architecturally foreclosed.

\textbf{Jurisdictional authority} is undermined by the borderless distribution of decentralized networks \citep{perloffgiles2018transnational, svantesson2004characteristics}. Activity can migrate to more permissive jurisdictions when governance pressure is applied, as Bitcoin mining did following China's 2021 ban \citep{thorn2021examining}.

\textbf{Responsiveness} fails most fundamentally. DeAI can break this condition in two ways: there may be no human principal to receive the address, or an identifiable principal may lack causal control over the running system. In either case, the failure is not one of deterrence or comprehension, but of effective purchase on the system itself.

\subsection{The Addressability Failure}

The failure of these presuppositions shows that applying governance through normative address to DeAI is a category error. The problem is not only that the addressee may be absent, but that the chain connecting addressee to system may be severed even when an addressee exists.

This failure has practical consequences. Prosecuting individuals associated with DeAI systems may punish those individuals without affecting system behavior, while pre-deployment and compliance requirements fail where no identifiable deployer exists. Law has long struggled to keep pace with technological change \citep{bennettmoses2007recurring}, but DeAI marks a qualitative escalation: the problem is no longer merely one of timing or jurisdiction, but of whether normative governance can reach its object at all.

Recent proposals attempt to repair this broken chain through legal-entity workarounds. \citet{arbel2026how}, confronting what they call the individuation problem---that AIs ``lack bodies'' and ``can copy, split, merge, swarm, and vanish at will''---distinguish thin identification from thick identification and propose the Algorithmic Corporation (A-corp), a human-owned legal entity operated by AIs. This is a valuable response to the accountability gap because it reconstructs an addressable principal. But it does not resolve the incapacitation gap: sanctioning the human owners of an A-corp does not by itself halt agents whose continued operation is secured by open-weight models, permissionless compute, and TEE-resident execution. Legal-personhood proposals repair addressability without restoring causal control. The A-corp can also be understood through the lens of \citeauthor{elish2019moral}'s \citeyearpar{elish2019moral} ``moral crumple zone'': just as the nearest human operator in a semi-automated system absorbs blame despite limited control, the human owners of an A-corp may absorb legal liability for an autonomous system they cannot actually alter or terminate. In economic terms, the governance vacuum describes the collapse of the principal-agent relationship \citep{jensen1976theory}: DeAI produces a condition of agency without a principal---an ``orphaned agent'' whose operation persists without any party capable of performing the monitoring, sanctioning, or redirecting functions that the principal-agent framework presupposes. Where centralized AI creates information asymmetries between principal and agent, DeAI eliminates the principal altogether.

Recognizing this category error does not mean DeAI is ungovernable in any absolute sense. It means governance must operate through a different modality: one that does not depend on an addressable agent, or on that agent's ability to alter the running system. What is needed is ``regulation by design,'' where governance requirements are enforced architecturally rather than through normative address \citep{almada2023regulation}. This is the case for protocol-based governance through architectural constraint.

\section{Governance Through Architectural Constraint: The Case for Protocol}
\label{sec:protocol}

\subsection{Lessig's Four Modalities and the Primacy of Architecture}

\citet{reidenberg1998lex} first argued that technology itself formulates policy rules---a ``Lex Informatica'' in which system design choices impose regulatory constraints functionally equivalent to legal rules. Lawrence Lessig's foundational framework generalized this insight, identifying four modalities through which behavior is regulated: law, social norms, markets, and architecture (or code) \citep{lessig1999code}. This was extended to blockchain governance by \citet{defilippi2018blockchain}, who argue that ``code is law'' takes on literal force when smart contracts enforce rules without institutional intermediation---a condition they term \emph{lex cryptographia}. \citet{defilippi2024ruleofcode} update this analysis, arguing that blockchain governance operates through a hybrid of code-based enforcement and community-driven norm-setting that cannot be reduced to either modality alone. Law regulates through the threat of sanction. Social norms regulate through the pressure of community disapproval. Markets regulate through the price mechanism. Architecture regulates by structuring the environment in which action occurs, making certain actions possible, impossible, easy, or difficult.

Lessig's key insight is that architecture is not merely one regulatory modality among four; in cyberspace, it is the most powerful and least visible, structuring the environment in which the other three operate. A highway's physical design constrains driving behavior more effectively than speed limit signs. A building's architecture determines who can access which spaces more reliably than ``authorized personnel only'' notices. Code, in the digital context, determines what users can and cannot do more powerfully than terms of service.

For DeAI, architecture---specifically, the protocols that govern how agents interact with blockchain infrastructure, acquire resources, and execute computations---is the only modality with direct regulatory purchase. Law cannot reach pseudonymous actors across multiple jurisdictions. Social norms cannot influence entities without social identity or reputation. Markets can shape agent behavior through incentive structures, but only if those incentive structures are encoded in the protocol. Protocol is the regulatory modality that remains effective when the other three fail. This is the pragmatic argument for protocol governance. But there is a deeper, ethical argument as well.

\subsection{Protocol as Architectural Constraint: The Ethical Argument}

The distinction between normative address and architectural constraint does not map neatly onto a temporal axis (e.g., \emph{ex post} versus \emph{ex ante}) because, as we have shown, normative address includes a full range of \emph{ex ante} instruments. Rather, the distinction concerns what the governance mechanism \emph{requires of the governed entity}. Normative address requires a comprehending, responsive agent who can receive communications and modify behavior accordingly. Architectural constraint requires no such agent. We argue that governance through protocols of architectural constraint is ethically appropriate for governing entities that lack addressability, for three reasons.

\emph{First}, architectural constraint does not presuppose addressability in the governed entity. A median barrier works regardless of the driver's intentions, knowledge of traffic laws, or susceptibility to deterrence. Similarly, protocol-level constraints on DeAI agents---such as cryptographic verification requirements, resource usage limits, and mandatory transparency interfaces---operate regardless of whether the agent can ``understand'' or ``intend'' compliance. This is not a bug but a feature: governance should match the ontological properties of the governed. Crucially, architectural constraint does not depend either on an identifiable principal or on that principal's ability to alter the system. It operates at the point of execution, on the substrate that hosts the action itself.

\emph{Second}, architectural constraint places the ethical burden where it belongs---on the designers of the governance architecture rather than on the governed entities. When governance operates through protocol, the moral questions shift from ``how do we communicate expectations to non-addressable entities?'' to ``what values should be embedded in the architecture?'' and ``who has the legitimate authority to make these design decisions?'' As \citet{winner1980artifacts} demonstrated, technical artifacts are never politically neutral: design choices embed specific forms of power and authority, whether intentionally or not. The question of what values to embed in protocol architecture is therefore a question of institutional design with political stakes \citep{friedman2002vsd}. These are questions of human moral responsibility---of protocol designers, standards bodies, and governance communities---rather than questions about the addressability of AI systems.

\emph{Third}, architectural constraint is honest about the nature of the relationship between human governance and autonomous systems. Normative address applied to DeAI creates a fiction of accountability---the pretense that someone is receiving and complying with governance communications when, in fact, no one is, or no one whose compliance would suffice. Protocol governance acknowledges that control, if it is to exist at all, must be built into the architecture from the outset. It replaces the governance fiction with a governance reality, even if that reality is more limited than the fiction it replaces.

This logic is already reflected in AI engineering practice. The emergence of ``harness engineering'' \citep{openai2026harness} embodies the insight that ``Agent = Model + Harness'' \citep{greyling2026harness}: when an agent misbehaves, the response is not to instruct the model (normative address) but to modify the harness that makes misbehavior architecturally impossible. Our argument extends this from individual agents to entire ecosystems: if harnesses govern individual agents, protocols govern the agentic web. Protocols are ``hard harnesses'' encoded in the substrate itself. But the analogy also reveals the stakes: a protocol governing a global decentralized network is an act of constitutional design---what \citet{suzor2018digital} calls ``digital constitutionalism''---and, given the immutability of blockchain-based protocols, may be extremely difficult to revise. The values embedded in these protocols are not preferences to be iterated upon in a product cycle but the foundational constraints that shape the possibility space for an ecosystem.

The transition from normative address to architectural constraint also raises a problem of authority. In conventional governance, architectural constraints derive legitimacy from a prior layer of normative address: building codes authorize fire exits; GDPR and the EU AI Act authorize privacy- and safety-by-design requirements. In DeAI, that authorizing chain breaks down. When policy fails to find an addressable subject, protocol may still constrain behavior, but its democratic authorization becomes unclear. This is not a secondary implementation detail but the central ethical challenge of protocol governance: the enforcer may persist after the commander disappears.

\subsubsection{From Authorization to Constitution: The Scaffolding of Protocol}

The shift to protocol-based governance does not imply the disappearance of policy, but its migration upstream. Critics of ``regulation by design'' \citep{almada2023regulation, hildebrandt2015smart, yeung2017hypernudge, yeung2018algorithmic} rightly note that protocols are never institutionally bare: constraining TEE manufacturers, L1 designers, or hardware vendors still requires institutional mechanisms. Even Bitcoin depends on a normative layer of BIPs, client maintainers, and social consensus. The shift from policy to protocol is therefore not the elimination of governance, but a transformation in how governance operates.

This institutional scaffolding is not merely traditional regulation under a different name. While policy may migrate upstream, the mode of governance changes in function as well as location. The distinction is not between the presence or absence of institutions, but between two qualitatively different modes of institutional intervention. Drawing on \citeauthor{searle1995construction}'s \citeyearpar{searle1995construction} distinction between regulative and constitutive rules, these modes can be categorized as \emph{regulative governance} and \emph{constitutive governance}.

\textbf{Regulative governance} addresses pre-existing actors and prescribes how they ought to behave: developers must conduct conformity assessments, deployers must implement risk management systems, operators must comply with data protection mandates. The rule presupposes the activity it regulates and threatens consequences for violation. Enforcement is \emph{ex post} in its essential structure even when its trigger conditions are \emph{ex ante}: someone must have failed to comply for the rule to bite, and the bite takes the form of fines, audits, criminal sanctions, or market exclusion imposed on an addressable principal. GDPR, the EU AI Act, and the NIST AI RMF are paradigmatic instances of regulative governance, even when they include ``by design'' obligations, because the obligations are ultimately addressed to persons.

\textbf{Constitutive governance}, by contrast, shapes the conditions of possibility for action through institutional scaffolding. It does not tell agents what to do but determines what counts as an action within a given system at all. As \citet{schauer2021regulative} argues, constitutive rules carry a ``regulative overhang'': by defining the official way of doing things, they make alternatives less eligible, less available, or less permitted---so that constituting what counts as a valid transaction simultaneously regulates which behaviors are possible. A protocol-level requirement that on-chain agents present cryptographically verifiable identifiers in order to access compute is constitutive in this sense: it does not threaten unauthorized agents with sanction but renders unauthorized action substrate-impossible. The institutional work happens upstream of deployment (e.g., in standard-setting bodies, in client implementations, in hardware certification pipelines, in L1 social consensus) and the work consists in writing the rules of the game rather than penalizing players who break them. This is institutional scaffolding, but of a constitutive rather than regulative kind: it creates the conditions under which action becomes possible in the first place.

Three contrasts crystallize the distinction. First, the \emph{locus}: regulative governance enforces against a principal at the moment of violation; constitutive governance enforces at the moment of execution, rendering non-compliant action impossible rather than punishable. Second, the \emph{temporality}: constitutive governance must be in place before the system exists---invoking the Collingridge dilemma \citep{collingridge1980social}. Protocol governance relocates this dilemma: the burden of foresight shifts from regulators to protocol designers. Constitutive governance forecloses certain futures rather than bending existing trajectories, making its political stakes correspondingly higher. Third, the \emph{addressee}: regulative governance addresses agents who use a technology; constitutive governance addresses the substrate-builders who determine what kinds of agents can exist. The governance vacuum at the agent level does not entail a governance vacuum at the substrate level. Address remains possible at the upstream layer precisely because it has failed at the downstream one. Protocols are \emph{ex ante}, self-enforcing, architectural, and constitutive of interaction---deviation is either impossible or tantamount to exit from coordination. Policies are \emph{ex post}, sanction-backed, authoritative, and regulative of conduct---deviation is possible, occurs, and is then punished by a third party. Protocols presuppose a coordination space they help bring into being; policies presuppose subjects on whom sanction can land. Table~\ref{tab:protocol-policy} summarizes the distinction across four analytical dimensions with illustrative examples.

\begin{table}[ht]
\centering
\caption{Protocol (Constitutive) vs.\ Policy (Regulative): four analytical dimensions.}
\label{tab:protocol-policy}
\small
\begin{tabular}{@{}p{2.2cm}p{5.2cm}p{5.2cm}@{}}
\toprule
\textbf{Dimension} & \textbf{Protocol (Constitutive)} & \textbf{Policy (Regulative)} \\
\midrule
\emph{Temporal structure} &
\emph{Ex ante}: constraints are in place before action occurs &
\emph{Ex post}: rules are enforced after violation is detected \\[4pt]
\emph{Enforcement mechanism} &
Self-enforcing and inviolable: non-compliant action fails to execute (e.g., ERC-8183 escrow reverts without evaluator attestation) &
Sanction-backed and violable: deviation occurs and is then punished by a third party (e.g., GDPR fines imposed by a data protection authority) \\[4pt]
\emph{Source of authority} &
Architectural: embedded in the technical substrate (e.g., consensus rules, cryptographic verification requirements) &
Authoritative: issued by a sovereign or delegated body (e.g., EU AI Act obligations addressed to deployers) \\[4pt]
\emph{Semantic function} &
Constitutive of interaction: defines what counts as a valid action within the system (e.g., ERC-8004 identity registry constitutes agent-hood) &
Regulative of conduct: prescribes how pre-existing actors ought to behave (e.g., mandatory impact assessments for high-risk AI) \\
\bottomrule
\end{tabular}
\end{table}

This shift toward constitutive governance clarifies that the ``governance vacuum'' at the agent level is a relocation, rather than an elimination, of the political problem. By migrating governance upstream, the actors who control the technical chokepoints---TEE vendors, L1 consensus communities, core maintainers, and certification bodies---emerge as the new locus of sovereignty within the decentralized ecosystem. These entities often operate with less democratic accountability than the regulatory agencies they displace, and the technical complexity of their decisions frequently obscures their deep political stakes. This represents a refined iteration of Lessig's ``code is law'' insight: while code may function as law, the question of who authors that code, under what authorization, and to what ends, becomes the central political tension of the constitutive mode. Rather than falling into technocracy by accident, constitutive governance renders the ``technocracy question'' a foundational element of the governance architecture itself. Section~\ref{sec:ethical_conditions} develops the ethical conditions required to ensure that such exercises of constitutive power remain legitimate.

\subsection{Constitutive Governance in Practice: Early Protocol Experiments}
\label{sec:protocols_in_practice}

The distinction between regulative and constitutive governance is not merely analytical. Emerging protocol standards on the Ethereum blockchain represent early-stage attempts to instantiate constitutive governance for autonomous AI agents. Two draft standards are instructive, each targeting one of the two failure modes identified in Section~\ref{sec:layers}: ERC-8004 (Trustless Agents) addresses the accountability gap, and ERC-8183 (Agentic Commerce Protocol) addresses the incapacitation gap \citep{erc8004, erc8183}. Neither is mature or widely deployed; both are draft proposals under community review. Their value for our argument lies not in their adoption but in the governance logic they embody---they illustrate what constitutive governance looks like when translated from theory into protocol design.

ERC-8004 responds to the accountability gap by constructing addressability at the protocol level \citep{erc8004}. Where conventional responses are regulative---mandate KYC, instruct deployers to register, or construct legal-entity workarounds such as \citeauthor{arbel2026how}'s Algorithmic Corporation---ERC-8004 takes a constitutive approach. It defines on-chain registries for identity, reputation, and validation that make these properties preconditions for ecosystem participation. An unregistered agent cannot accumulate reputation, cannot be validated, and cannot be discovered. The protocol does not instruct anyone to identify themselves; it renders unidentified action substrate-impossible. In Searle's terms, it constitutes what it means to be an agent within the system rather than regulating agents who already exist within it. This also offers a partial response to the detectability problem: where TEE-protected computation forecloses direct observation, the Validation Registry enables cryptographic verification of outputs without access to the computation itself---relocating transparency from the process to the protocol.

ERC-8183 responds to the incapacitation gap \citep{erc8183}. Its smart-contract state machine enforces a job lifecycle in which no value flows without evaluator attestation---unlike Tornado Cash, which is a protocol of pure execution with no embedded governance leverage. Agents can transact entirely without human intervention, yet governance constraints are enforced at every state transition. Optional hook contracts extend this logic: a pre-funding hook that reverts when a reputation threshold is unmet does not \emph{report} a violation but makes the non-compliant action \emph{fail to execute}---the constitutive mode rendered in code.

The two protocols compose: identity gates reputation, reputation gates commerce, and commerce requires attestation. At no point does this chain depend on identifying a human principal. These are, however, early experiments with significant limitations: they govern only agents that enter the protocol's state space, and they are vulnerable to enforcement migration. The question of who designs these protocols, under what authorization, and with what legitimacy is precisely the question the following section addresses.

\subsection{The Ethical Conditions for Protocol Governance: Addressing the Legitimacy Crisis}
\label{sec:ethical_conditions}

If protocol governance is to be normatively defensible rather than merely effective, it must satisfy several ethical conditions. These respond directly to the legitimacy problem created when architectural constraint persists after the ordinary chain of policy authorization has broken down. We identify four conditions.

\textbf{Legitimacy.} Protocol governance exercises power by determining what agents can and cannot do. In conventional governance, this power is authorized by democratic institutions: building codes authorize fire exits; GDPR authorizes privacy-by-design. For DeAI, this authorization chain is severed \citep{rawls1971theory, habermas1996between}. DAO-based governance structures offer one model \citep{wright2015decentralized}, but face well-documented challenges: plutocratic voting power, low participation, and exclusion of non-token-holders \citep{barbereau2023decentralised}. \citeauthor{ostrom1990governing}'s \citeyearpar{ostrom1990governing} design principles for commons governance and her concept of polycentric governance offer alternative frameworks, but only if they can resist the recentralization dynamics that empirical studies have documented \citep{rong2025governing}.

\textbf{Contestability.} Values embedded in protocols must be contestable---such that they are subject to challenge, revision, and override through legitimate processes. The immutability that makes blockchain-based protocols resistant to unilateral interference also makes them resistant to democratic revision. This is a feature when it protects against authoritarian censorship; it is a problem when it prevents the correction of unjust or harmful design choices. Protocol governance must include mechanisms for structured contestation---such as upgrade processes, governance forks, and sunset clauses---that balance stability against the capacity for moral learning and correction.

\textbf{Transparency.} If protocol governance replaces the observability of agent behavior (which radical opacity forecloses) with the observability of governance architecture, then the protocols themselves must be transparent. This means not merely open-source code, but human-readable documentation of the values, constraints, and trade-offs embedded in the architecture. \citeauthor{rahwan2018society}'s \citeyearpar{rahwan2018society} insight that transparency must concern the external behavior of systems, not merely their source code, applies with equal force to governance protocols: stakeholders need to understand what the protocol does, not merely how it is coded.

\textbf{Non-domination.} Drawing on republican political theory \citep{pettit1997republicanism}, protocol governance must be designed to prevent domination---that is, the capacity of any actor or group to exercise arbitrary power over others through control of the governance architecture. \citet{hoeksema2023domination} argues that even radical republican accounts are needed for digital platforms because individual-agent framings miss structural domination; DeAI, where no addressable principal exists, represents the limit case of such structural domination through architecture. This includes preventing capture by protocol designers, wealthy token-holders, or powerful node operators. The history of blockchain governance---including the recentralization dynamics documented in empirical studies of DAOs and decentralized systems \citep{rong2025governing}---demonstrates that decentralized architectures are not inherently immune to power concentration.

\section{Conclusion: The Ethics of Architectural Governance}
\label{sec:conclusion}

DeAI does not simply add a new topic to AI ethics; it unsettles the assumptions on which existing governance frameworks depend. When there is no reliable addressee for governance, or when addressing that addressee has no effect on the running system, governance through normative address loses purchase. DeAI makes both failures possible: an accountability gap, in which no addressable principal can be identified, and an incapacitation gap, in which even an identifiable principal cannot alter the system. Under these conditions, governance shifts from the agent to the architecture itself.

We have argued that this requires a shift from normative address to architectural constraint---an operationalization of what \citet{floridi2013distributed} calls ``infraethics'': the framework of background conditions that makes ethical action possible. This is not a retreat from institutions, but a relocation of governance upstream: from regulating agents within a system to shaping the substrates within which agents operate. Protocol-based architectural constraint is ethically appropriate here because it does not depend on the principal-agent chain that DeAI destabilizes. But this shift also generates a legitimacy problem. In conventional governance, architectural constraints are authorized by a prior policy layer. In DeAI, that chain is fractured. Protocol governance must therefore develop alternative sources of legitimacy or risk becoming an unaccountable exercise of technocratic power.

No existing proposal fully resolves this problem. Zero-knowledge verification, DAO governance, AI identification systems, and society-in-the-loop frameworks each address some of the relevant ethical conditions---of legitimacy, contestability, transparency, and non-domination---while falling short on others \citep{chan2024visibility, chan2024ids, wright2015decentralized, barbereau2023decentralised, rahwan2018society}. A defensible governance regime will likely require layered integration: protocol-level constraints embedded within socio-technical structures that render those constraints legitimate, transparent, and contestable.

Two risks are especially salient. The first is \emph{technocratic capture}: if governance is embedded in architecture, then protocol designers, standards bodies, and core developers may acquire an outsized and insufficiently accountable form of power \citep{denardis2014global}. The second is \emph{enforcement migration}: agents may evade governance by moving to more permissive chains. Yet this coordination problem is more tractable than the one that defeats normative governance, because it shifts attention upstream to a smaller set of addressable substrate-builders rather than downstream to absent or anonymous principals.

The question, then, is not whether DeAI can be perfectly governed, but whether governance architectures can be made ethically defensible. The governance vacuum created by DeAI will not be solved simply by finding new entities to address. It will be addressed, if at all, by building governance into decentralized architectures and by reconstructing forms of authorization that make those architectures legitimate. That is the central ethical and political challenge of governing AI in a decentralized world.

\bibliography{references}

\end{document}